\documentclass[conference]{IEEEtran}
\IEEEoverridecommandlockouts
\usepackage{cite}
\usepackage{amsmath,amssymb,amsfonts}
\usepackage{algorithmic}
\usepackage{graphicx}
\usepackage{textcomp}
\usepackage{xcolor}
\usepackage{stfloats}
\usepackage{mathrsfs}
\usepackage{balance}
\usepackage{svg}

\usepackage[ruled,vlined]{algorithm2e}

\allowdisplaybreaks
\begin{document}

\title{Reduced-latency DL-based Fractional Channel Estimation in OTFS Receivers
}

\author{\IEEEauthorblockN{Mauro Marchese\IEEEauthorrefmark{1}, Henk Wymeersch\IEEEauthorrefmark{2}, Paolo Spallaccini\IEEEauthorrefmark{3}, Stefano Chinnici\IEEEauthorrefmark{3}, Pietro Savazzi\IEEEauthorrefmark{1}\IEEEauthorrefmark{4}}
\IEEEauthorblockA{\IEEEauthorrefmark{1}University of Pavia, Italy, \IEEEauthorrefmark{3}HCL Software, Vimodrone, Milan, Italy, \IEEEauthorrefmark{2}Chalmers University of Technology, Sweden, \\\IEEEauthorrefmark{4}CNIT Consorzio Nazionale Interuniversitario per le Telecomunicazioni, Pavia, Italy\\
E-mail: mauro.marchese01@universitadipavia.it}
}

\maketitle

\begin{abstract}
In this work, we propose a deep learning (DL)-based approach that integrates a state-of-the-art algorithm with a time-frequency (TF) learning framework to minimize overall latency. Meeting the stringent latency requirements of 6G orthogonal time-frequency space (OTFS) systems necessitates low-latency designs. The performance of the proposed approach is evaluated under challenging conditions: low delay and Doppler resolutions caused by limited time and frequency resources, and significant interpath interference (IPI) due to poor separability of propagation paths in the delay-Doppler (DD) domain. Simulation results demonstrate that the proposed method achieves high estimation accuracy while reducing latency by approximately 55\% during the maximization process. However, a performance trade-off is observed, with a maximum loss of 3 dB at high pilot SNR values.
\end{abstract}

\begin{IEEEkeywords}
Deep learning (DL), interpath interference (IPI), channel estimation, fractional channel parameters.
\end{IEEEkeywords}

\section{Introduction}
Channel estimation in orthogonal time-frequency space (OTFS) \cite{Hadani2017,HadaniMonk2018,Viterbo2022} systems plays a crucial role in future 6G networks, as data detection heavily relies on accurate channel state information (CSI) at the receiver. In OTFS systems, the channel parameters often assume fractional delay and Doppler values when delay and Doppler resolutions are insufficient. This leads to the spreading of received pilot replicas into adjacent bins, causing interpath interference (IPI) \cite{Muppaneni2023}. To address this, multigrid algorithms that account for fractional delays and Doppler shifts must be employed to improve estimation accuracy. Additionally, refinement steps are needed to enhance detection capabilities and eliminate false paths introduced by IPI \cite{marchese2024}. Thus, an increase in complexity is unavoidable to guarantee high estimation quality. Therefore, novel solutions should be developed to reduce latency due to receiver processing when such computationally demanding algorithms must be used.

To mitigate the impact of IPI, the progressive interpath interference cancellation (P-IPIC) algorithm \cite{marchese2024} was proposed as a low-latency alternative to the state-of-the-art delay-Doppler interpath interference cancellation (DDIPIC) algorithm \cite{Muppaneni2023}. P-IPIC achieves better performance than DDIPIC, with gains of approximately $1-2$ dB in high-IPI scenarios while significantly reducing latency. The latency reduction stems from (i) a single global refinement step in P-IPIC, compared to multiple refinements required by DDIPIC which cancels IPI through these refinement procedures; (ii) less complexity needed for cost function evaluation since the P-IPIC cancels IPI by computing a residue vector and estimates channel parameters by maximizing a residue-based cost fucntion \cite{marchese2024}. While other low-complexity methods, such as the modified maximum likelihood estimator (M-MLE) \cite{Khan2021}, have been explored, DDIPIC is shown to outperform these methods at the expense of higher computational complexity \cite{Muppaneni2023}.

Deep learning has emerged as a transformative tool in wireless communications \cite{Dai2020}, providing solutions to challenges in 6G technologies by enhancing performance and reducing complexity. For example, \cite{Guo2023} demonstrates that a deep neural network (DNN) can directly estimate channel parameters from received frames, outperforming conventional threshold-based methods \cite{Raviteja2019}. Similarly, \cite{Gao2024} highlights the suitability of deep learning for feature extraction in the delay-Doppler domain. In \cite{Mattu2024}, a DNN-based approach is proposed to reduce the latency of the DDIPIC algorithm \cite{Muppaneni2023}. This approach approximates the relationship between delay-Doppler pairs and corresponding columns of a constituent delay-Doppler parameter matrix (CDDPM), with training in the time-frequency domain achieving superior performance due to the narrower value range of unvectorized CDDPM columns.

In this work, the goal is to develop a reduced-latency algorithm for channel estimation in high-IPI regime where higher complexity is needed to counteract the effects if IPI. The P-IPIC algorithm from \cite{marchese2024} is considered as the basline method as it is a state-of-the art algorithm for channel estimaiton in high IPI regimes. To achieve this latency reduction, the computation of the cost function, which is inevitably evaluated multiple times during the estimation procedure, should be simplified. To do so, the deep learning approach from \cite{Mattu2024} is adapted and combined with the recently proposed P-IPIC algorithm in \cite{marchese2024} that leverages a global refinement to reduce IPI. Therefore, two neural networks are designed to predict the time-frequency CDDPM required for residue-based cost function computation. Simulation results demonstrate a significant reduction in latency, approximately halving the time required for cost function maximization, while maintaining high estimation accuracy. The proposed DL-based P-IPIC is therefore shown to be a good candidate as low-latency channel estimation algorithm for high-IPI scenarios.

\textit{Notation}:  $\mathbf{X}$ is a matrix, $ \mathbf{x}$ is a vector, and $x$ is a scalar. $ X[m,n]$ represents the $(m,n)$-th element of the matrix $\mathbf{X}$, and $\hat{x}$ denotes the estimate of $x$. $\|\mathbf{X}\|_F$, $\mathbf{X}^T$, $\mathbf{X}^H$, and $\mathbf{X}^{-1}$ represent the Frobenius norm, transpose, Hermitian (conjugate transpose), and inverse of $\mathbf{X}$, respectively. $\mathbf{I}$ is the identity matrix, $\|\mathbf{x}\|$ is the norm of the vector $\mathbf{x}$, and $|x|$ denotes the absolute value of scalar $x$. The operators $ \text{vec}(\mathbf{X})$ and $\text{vec}_{M,N}^{-1}(\mathbf{x})$ represent the vectorization of matrix $\mathbf{X}$ and the reshaping of vector $\mathbf{x}$ back into a $M\times N$ matrix, respectively. $\mathbb{E}\{\cdot\}$ denotes the expected value and $\mathcal{CN}(0, \sigma^2 \mathbf{I})$ represents a complex Gaussian random variable with zero mean and covariance matrix $\sigma^2 \mathbf{I}$. Finally, $\mathbb{C}$ denotes the set of complex numbers.

\section{OTFS Model and Channel Estimation}

\begin{figure*}[b]
    \begin{align*}
        \Upsilon_i[k'M+l',k''M+l'']&=\frac{e^{-j2\pi\tau_i\nu_i}}{MN}\sum_{n=0}^{N-1}\sum_{m=0}^{M-1}f_{k'',l'}(m)e^{j2\pi\big(\frac{m}{M}\big(l'-l''-\frac{M\tau_i}{T}\big)-\frac{n}{N}\big(k'-k''-\frac{N\nu_i}{\Delta f}\big)\big)}, \\
        f_{k'',l'}(m)& =\sum_{s=-m}^{M-1-m}e^{j2\pi s\frac{l'}{M}}\Big[\Big(1-\frac{\tau_i}{T}\Big)e^{j\pi\big(1+\frac{\tau_i}{T}\big)\big(\frac{\nu_i}{\Delta f}-s\big)}\textnormal{sinc}\Big(\Big(1-\frac{\tau_i}{T}\Big)\Big(\frac{\nu_i}{\Delta f}-s\Big)\Big)+ \\
        & \quad + \frac{\tau_i}{T}e^{-j2\pi\frac{k''}{N}}e^{j\pi\big(\frac{\tau_i}{T}\big)\big(\frac{\nu_i}{\Delta f}-s\big)}\textnormal{sinc}\Big(\tau_i\big(\nu_i-s\Delta f\big)\Big)\Big].
    \end{align*}
\end{figure*}

\subsection{System Model}
In this section, both pilot and observation models for fractional channel estimation are presented in OTFS systems with rectangular pulse shaping. In the following, an OTFS system with $M$ subcarriers with spacing $\Delta f$ and $N$ time slots of duration $T$ is considered.

\subsubsection{Pilot model}  
The single-antenna transmitter sends a pilot symbol in the delay-Doppler (DD) domain. The DD pilot frame is given by
\begin{align} \label{eq:1}
X[m,n]=
    \begin{cases}
    \sqrt{E_p} & m=m_p,~n=n_p, \\
    0 & \textnormal{otherwise},
    \end{cases}
\end{align}
where ($m_p,n_p$) is the delay-Doppler cell in which the pilot is sent and $E_p$ is the energy of the transmitted pilot signal. 
\subsubsection{Observation model}
the received vectorized pilot frame is given by
\begin{align} \label{eq:2}
\textbf{y} =\textbf{H}_{dd}\textnormal{vec}(\textbf{X})+\textbf{n},
\end{align}
where $\textbf{n}\sim\mathcal{CN}(0,\sigma^2\textbf{I})\in\mathbb{C}^{MN\times 1}$ is the AWGN noise vector and $\textbf{H}_{dd}\in\mathbb{C}^{MN\times MN}$ is the delay-Doppler domain channel matrix given by
\begin{equation} \label{eq:3}
\textbf{H}_{dd}=\sum_{i=1}^P\alpha_i\boldsymbol{\Upsilon}_i(\tau_i,\nu_i),
\end{equation}
where $P$ is the number of propagation paths, $\alpha_i$ is the complex gain of the $i$-th path and $\tau_i$ and $\nu_i$ are the propagation delay and the Doppler shift associated with the $i$-th path, respectively. Moreover, $\boldsymbol{\Upsilon}_i(\tau_i,\nu_i)\in\mathbb{C}^{MN\times MN}$ is a matrix that captures the effect of the propagation delay and the Doppler shift of the $i$-th path and for $l',l''=0,1,..,M-1$,~$k',k''=0,1,...,N-1$ is computed according to equations reported at the bottom of the page (see also \cite{marchese2024,Muppaneni2023,Khan2021,Gaudio2020,Mattu2024}). Channel estimation is performed considering the input-output relation rewritten as
\begin{equation} \label{eq:4}
\textbf{y}=\textbf{R}\big(\boldsymbol{\tau},\boldsymbol{\nu}\big)\boldsymbol{\alpha}+\textbf{n},
\end{equation}
where $\boldsymbol{\alpha}=[\alpha_1~\alpha_2~...~\alpha_P]^T\in\mathbb{C}^{P\times 1}$ is the channel gains vector and $\textbf{R}\big(\boldsymbol{\tau},\boldsymbol{\nu}\big)=[\textbf{r}_1~\textbf{r}_2~...~\textbf{r}_P]\in\mathbb{C}^{MN\times P}$ is the CDDPM matrix. Each column of $\textbf{R}\big(\boldsymbol{\tau},\boldsymbol{\nu}\big)$ is related to a different path and is given by
\begin{equation} \label{eq:5}
\textbf{r}_i(\tau_i,\nu_i)=\boldsymbol{\Upsilon}_i(\tau_i,\nu_i)\textnormal{vec}(\textbf{X}).
\end{equation} 

\begin{figure*}[t]
\centering
    \includegraphics[width=\textwidth, trim=50 200 50 120, clip]{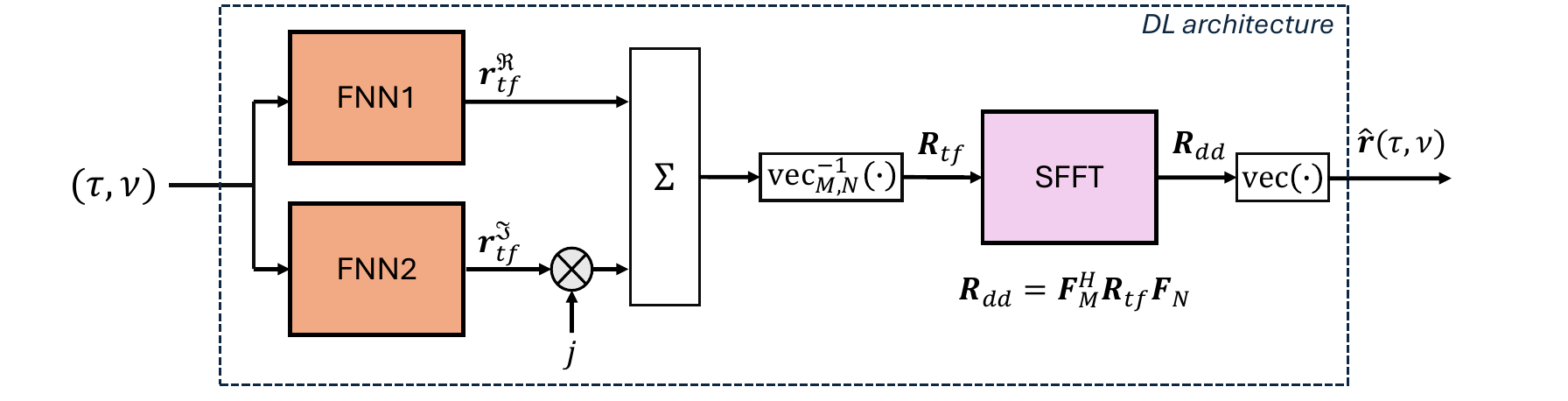}
    \caption{The architecture of the TF learning approach: a delay-Doppler pair is mapped to the corresponding real and imaginary parts of the TF CDDPM column. After that proper reshaping and conversion to the DD domain is applied to obtain the desired CDDPM column.}\label{fig:FNNarchitecture}
\end{figure*}

\subsection{The P-IPIC Algorithm}
In this section the P-IPIC algorithm proposed in \cite{marchese2024} is summarized. This channel estimation algorithm operates in two distinct phases.

    \subsubsection{Search phase} The algorithm iteratively searches for possible paths maximizing a residue-based cost function given by
    \begin{equation} \label{eq:6}
    \Phi_e\big(\textbf{r}(\tau, \nu)\big)=\frac{|\textbf{r}^H(\tau, \nu)\textbf{e}^{(i-1)}|^2}{\|\textbf{r}(\tau, \nu)\|^2},
    \end{equation} 
    where the residue vector at the $i$-th iteration is given as
    \begin{equation} \label{eq:7}
    \textbf{e}^{(i)}=\textbf{y}-\textbf{R}(\hat{\boldsymbol{\tau}},\hat{\boldsymbol{\nu}})\hat{\boldsymbol{\alpha}}.
    \end{equation}
   During this phase, the propagation delay and Doppler shift of the $i$-th path are estimated by maximizing $\Phi_e\big(\textbf{r}(\tau, \nu)\big)$ over a search area comprising delays and Doppler shifts that are multiples of the DD resolutions. It can be observed that the delay resolution is $\Delta\tau = \frac{1}{M\Delta f}$, and the Doppler resolution is $\Delta\nu = \frac{1}{NT}$. In this way, a coarse estimate is obtained.  

   Afterward, an iterative fine estimation procedure is performed to refine the estimated channel parameters. This is achieved by maximizing the residue-based cost function over a search area comprising fractional DD parameters centered at the estimate obtained in the previous step. The search area is narrowed down during each iteration, reducing the DD spacings until the desired accuracy is achieved or a maximum number of iterations is reached. Further details can be found in Section \ref{DLPIPIC} and in \cite{marchese2024}.
    Once the DD pair associated with the $i$-th path is estimated, the channel gains vector is updated using the regularized least squares (RLS) formula
    \begin{equation} \label{eq:8}
    \hat{\boldsymbol{\alpha}} = \Big(\textbf{R}^{H}(\hat{\boldsymbol{\tau}},\hat{\boldsymbol{\nu}})\textbf{R}(\hat{\boldsymbol{\tau}},\hat{\boldsymbol{\nu}})+\lambda\textbf{I}\Big)^{-1} \textbf{R}^{H}(\hat{\boldsymbol{\tau}},\hat{\boldsymbol{\nu}})\textbf{y},
    \end{equation}
    where $\lambda$ is a sufficiently small number.
    After that, once the residue vector is obtained, if $\|\textbf{e}^{(i)}\| < \epsilon$, where $\epsilon$ is the convergence tolerance parameter, the \textit{search phase} ends and a number of paths equal to $\hat{P} = i$ is detected. Otherwise, a new iteration is run to search for a new path. The threshold for the stopping criterion is chosen, assuming a Gaussian stationary noise source, as $\epsilon = 3\sqrt{MN\sigma^2}$.

    \subsubsection{Refinement phase} 
    During this second phase, the algorithm iteratively refines the estimated paths by performing a coarse and fine estimate, during which the DD pair associated with the $i$-th path is refined by maximizing the regularized observation-based cost function

    \begin{equation} \label{eq:9}
   \Phi_y\big(\textbf{R}\big)=\textbf{y}^H\textbf{R}\Big(\textbf{R}^H\textbf{R}+\lambda\textbf{I}\Big)^{-1}\textbf{R}^H\textbf{y},
    \end{equation} 
    where the dependence on the DD vectors is omitted for simplicity.
    This is done since $ \Phi_y\big(\textbf{R}\big)$ takes into account all the estimated paths. During the refinement phase, the residue vector is computed in each iteration, and the stopping criterion is checked. If the stopping criterion is met before all the paths are refined, the remaining unrefined paths are declared false alarms and discarded. Thus, during this phase, a number of paths equal to $\hat{\hat{P}} \leq \hat{P}$ is detected.

\section{DL-based Approach}
 During the cost function maximization, the computational complexity of the P-IPIC algorithm is dominated by the computation of the CDDPM column required to evaluate the residue-based cost function in \eqref{eq:5}. A reduced-latency approach is to replace the direct computation of \eqref{eq:5} with a DL architecture that learns and predicts the desired CDDPM column \cite{Mattu2024}. In this case, a performance loss is unavoidable due to the limited capabilities of a neural network to learn the mathematical relation between a DD pair and the corresponding CDDPM column. Moreover, the DL architecture outputs the TF version of the desired CDDPM column to facilitate training \cite{Mattu2024}.

\subsection{The Architecture}
Fig.~\ref{fig:FNNarchitecture} shows the adopted DL architecture. It consists of two distinct fully connected feedforward neural networks (FNNs) that learn the real and imaginary parts of the TF CDDPM column, respectively. In particular, the FNN is characterized by:
\begin{itemize}
    \item \textit{Two inputs}: the network receives as input a normalized DD pair. Normalization is performed to ensure that the delay values are in the range $[0,1]$ and the Doppler shift values are in the range $[-1,1]$.

    \item \textit{Concatenation layer}: this layer is used to concatenate the two inputs.

    \item \textit{Fully connected layers}: a certain number of fully connected layers are used. The activation function is ReLU, except for the output layer, which adopts a linear activation function.

    \item \textit{Output layer}: it provides the final output of the network. Its dimension is $MN$, as the TF CDDPM column must be provided as output.
\end{itemize}

As depicted in Fig.~\ref{fig:FNNarchitecture}, FNN1 learns the real part of the TF CDDPM column, denoted as $\textbf{r}^{\Re}_{tf}=\Re\{\textbf{r}_{tf}\}$, while FNN2 learns the imaginary part $\textbf{r}^{\Im}_{tf}=\Im\{\textbf{r}_{tf}\}$. The TF CDDPM column is given by the inverse symplectic finite Fourier transform (ISFFT) of the unvectorized CDDPM column $\textbf{R}_{dd}=\text{vec}_{M,N}^{-1}\big(\hat{\textbf{r}}\big(\tau,\nu)\big)$ as 
\begin{equation} \label{eq:10}
\textbf{r}_{tf}=\text{vec}\big(\textbf{R}_{tf}\big)=\text{vec}\big(\textbf{F}_{M}\textbf{R}_{dd}\textbf{F}_{N}^H\big),
\end{equation}
where $\textbf{F}_{N}=\frac{1}{\sqrt{N}}\big\{e^{-j2\pi\frac{mq}{N}}\big\}_{m,q=0}^{N-1}$ is a $N$-point discrete Fourier transform (DFT) matrix. Hermitian conjugate operation is applied to perform inverse DFT.

Hence, the two network outputs are combined and reshaped to obtain $\textbf{R}_{tf}$. Then, the SFFT is applied to obtain the DD matrix, which, once vectorized, provides the desired CDDPM column.

The number of hidden layers is 2. Increasing the number of hidden layers does not enhance performance but instead increases latency. The number of neurons per hidden layer is denoted as $L_1$ for the first hidden layer and $L_2$ for the second hidden layer. The input dimension of the network is 2, corresponding to a delay-Doppler pair. The last layer, with $MN$ neurons, is used to obtain the desired output dimension, and a linear activation function is used to span the real set.

\subsection{Offline Training}
The FNN-based architecture is trained to learn the desired relation. Training is performed using the backpropagation algorithm \cite{Haykin2009} and the Adam optimizer. The hyperparameters used during the training phase are reported in Table~\ref{table}. The learning rate is halved every 10 epochs. The objective function is the L1 loss function. Denoting $r_{tf}[n]$ the output of the $n$-th neuron in the output layer and $r_{tf}^d[n]$ the desired value for that output, the L1 loss function is defined as
\begin{equation} \label{eq:11}
\mathscr{E}=\sum_{n=1}^{MN}\Big|r_{tf}^d[n]-r_{tf}[n]\Big|.
\end{equation}
Training is carried out offline as described above and in the same way for the two networks. 

\begin{table}
\centering
\caption{Hyper-parameters for training.}
\normalsize{
\begin{tabular}{||c | c||} 
 \hline
Number of training samples & $200000$ \\ 
 \hline
Mini batch size & $1000$ \\  
 \hline
Number of epochs  & $100$ \\
 \hline 
Initial learning rate & $0.001$ \\
 \hline 
Learning rate drop factor  & $0.5$ \\
 \hline 
 Learning rate drop period  & $10$ \\
 \hline 
\end{tabular}
}
\label{table}
\end{table}

\subsection{DL-based P-IPIC}\label{DLPIPIC}
The DL-based P-IPIC channel estimation algorithm works as follows: during the cost function maximization, the required CDDPM column is obtained using FNN predictions. After that, before computing the channel gains vector, in order to maintain high estimation accuracy, the CDDPM column corresponding to the detected path is evaluated according to \eqref{eq:5} and stored. Hence, the maximum number of evaluations of \eqref{eq:5}, considering both \textit{search phase} and \textit{refinement phase}, is $2P_{\text{max}}$.  The algorithm flow is summarized in Algorithm~\ref{alg:1}. 

The DL-based P-IPIC requires the following parameters:
 $P_{\text{max}}$: maximum number of detectable paths; 
 $s_{\text{max}}$: maximum number of iterations during the fine estimation phase; 
    $\epsilon_{\tau},\epsilon_{\nu}$: convergence tolerance parameters for stopping criteria during fine estimate phase \cite{marchese2024,Muppaneni2023}; and 
    $m_{\tau},n_{\nu}$: parameters that are used to define the search space dimension along delay and Doppler axes during the fine estimation. In particular the number of grid points is $\big(2\lfloor{\frac{m_{\tau}}{2}}\rfloor+1\big)\big(2\lfloor{\frac{n_{\nu}}{2}}\rfloor+1\big)$.
Moreover, $I_{dd}$ indicates the integer DD grid used during the coarse estimate phase while $F_{dd}^{(s)}$ indicates the fractional DD grid used during the $s$-th iteration of the fine estimation phase. In particular, $F_{dd}^{(s)} = \Big\{W_{\tau}^{(s)}\Gamma+\hat{\tau}_{\text{fine}}^{(s-1)}\Big\} \times \Big\{W_{\nu}^{(s)}\Lambda+\hat{\nu}_{\text{fine}}^{(s-1)}\Big\}$  ($\times$ denots Cartesian product) where $\hat{\tau}_{\text{fine}}^{(s-1)}$,$\hat{\nu}_{\text{fine}}^{(s-1)}$ are the DD parameter estimates obtained in the previous iteration, $\Gamma=\{-\lfloor{\frac{m_{\tau}}{2}}\rfloor,...,0,...,\lfloor{\frac{m_{\tau}}{2}}\rfloor\}$ and $\Lambda=\{-\lfloor{\frac{n_{\nu}}{2}}\rfloor,...,0,...,\lfloor{\frac{n_{\nu}}{2}}\rfloor\}$ are the spanning sets, and $W_{\tau}^{(s)} = \frac{\Delta\tau}{m_{\tau}^{s-1}}$,$W_{\nu}^{(s)} = \frac{\Delta\nu}{n_{\nu}^{s-1}}$ are the fine delay and Doppler resolutions, respectively. It should be noted that, when $s=1$, $\hat{\tau}_{\text{fine}}^{(s-1)}=\hat{\tau}_{\text{coarse}}$, $\hat{\nu}_{\text{fine}}^{(s-1)}=\hat{\nu}_{\text{coarse}}$. In Algorithm~\ref{alg:1} the DD parameters of the search areas are fed as inputs to the DL architecture in Fig.~\ref{fig:FNNarchitecture} to obtain $\textbf{R}_{tf}=\text{vec}_{M,N}^{-1}\big(\text{FNN1}(\tau,\nu)+j\text{FNN2}(\tau,\nu))\big)$ and then the estimate of the corresponding CDDPM vector as $\hat{\textbf{r}}(\tau,\nu)=\text{vec(SFFT}(\textbf{R}_{tf}))$. The estimated CDDPM column $\hat{\textbf{r}}(\tau,\nu)$ is then used to evaluate the cost function during each phase.

\subsection{Latency Reduction}\label{LatRed}
The two FNNs must be designed properly in order to guarantee latency reduction by selecting a proper dimension for each layer. In particular, the complexity of the brute force evaluation of \eqref{eq:5} is of the order of $\mathcal{O}(N^3M^4)$. On the other hand, the latency of the FNNs, assuming $L_1, L_2 > MN$, is dominated by the second hidden layer and is approximated by $2L_1L_2$. With this in mind, the number of neurons per layer can be selected to ensure a latency gain. Specifically, the layer dimensions should satisfy $L_1L_2<\frac{N^3M^4}{2}$. In the following two cases are considered: $L_1=L_2$ and $L_2=4L_1$. Assuming $L_1=L_2$, the number of neurons per layer should be $L_1<\sqrt{\frac{N^3M^4}{2}}$. Conversely, when $L_2=4L_1$ the dimension of the first hidden layer should satisfy $L_1<\sqrt{\frac{N^3M^4}{8}}$. Moreover, in practical hardware implementation the use of FNNs facilitates the incorporation of parallelization, thus contributing to further reduction in latency.

\begin{algorithm}[t]
\caption{DL-based P-IPIC}\label{alg:1}
\KwIn{$P_{\max},\ \epsilon,\ s_{\max},\ \epsilon_{\tau},\ \epsilon_{\nu},\  m_{\tau},\ n_{\nu},\ \textbf{y}$}
\textbf{Initialize:} $\textbf{e}^{(0)}=\textbf{y}$\;
\textit{Search phase}:\newline
\For{$i = 1$ \KwTo $P_{\max}$}{
    Coarse estimate:\newline
    $\hat{\tau}_{\text{coarse}},\hat{\tau}_{\text{coarse}}=\arg\max_{(\tau,\nu)\in I_{dd}}  \Phi_e\big(\hat{\textbf{r}}(\tau,\nu)\big)$\;

    Iterative fine estimate ($s=1:s_{\max}$):\newline
    $\hat{\tau}_{\text{fine}}^{(s)},\hat{\nu}_{\text{fine}}^{(s)}=\arg\max_{(\tau,\nu)\in F_{dd}^{(s)}}  \Phi_e\big(\hat{\textbf{r}}(\tau,\nu)\big)$   $\hat{\tau}_i,\hat{\nu}_i=\hat{\tau}_{\text{fine}}^{(s)},\hat{\nu}_{\text{fine}}^{(s)}$\;

    Compute CDDPM column:
    $\textbf{r}_i(\hat{\tau}_i,\hat{\nu}_i)=\boldsymbol{\Upsilon}_i(\hat{\tau}_i,\hat{\nu}_i)\textnormal{vec}(\textbf{X})$\;

    Update channel gains vector:
    $\hat{\boldsymbol{\alpha}} = \Big(\textbf{R}^{H}(\hat{\boldsymbol{\tau}},\hat{\boldsymbol{\nu}})\textbf{R}(\hat{\boldsymbol{\tau}},\hat{\boldsymbol{\nu}})+\lambda\textbf{I}\Big)^{-1} \textbf{R}^{H}(\hat{\boldsymbol{\tau}},\hat{\boldsymbol{\nu}})\textbf{y}$\;

    Compute residue vector: \newline
    $\textbf{e}^{(i)}=\textbf{y}-\textbf{R}(\hat{\boldsymbol{\tau}},\hat{\boldsymbol{\nu}})\hat{\boldsymbol{\alpha}}$\;

    Check stopping criterion: \newline
    \If{$\|\textbf{e}^{(i)}\| < \epsilon$}{
            $\hat{P} = i$\;
            \textbf{break}\;
    }
}
\textit{Refinement phase} $(i_{\text{ref}} = 1: \hat{P})~\rightarrow\hat{\hat{P}}$\;

\textbf{Output:}~$\hat{\textbf{H}}_{dd}=\sum_{i=1}^{\hat{\hat{P}}}\hat{\alpha}_i\boldsymbol{\Upsilon}_i(\hat{\tau}_i,\hat{\nu}_i)$\;
\end{algorithm}

\section{Simulation Results}
In this section, the simulation results comparing the DL-based approach with the model-based approaches are presented. In the following, two different cases are considered: $L_1=L_2=8MN$ and $L_2=4L_1$ with $L_1=16MN$, where the dimension of the layers is selected according to the conditions reported in Section \ref{LatRed}. This analysis aims to investigate the learning properties of FNNs and the latency-performance trade-off.

\subsection{Simulation Parameters}
OTFS modulation with $M=N=16$ with subcarrier spacing of $\Delta f=\frac{1}{T}=25~\text{kHz}$ and carrier frequency $f_c=5.1~{\text{GHz}}$ is considered. Thus, delay and Doppler resolutions are 
$\tau_{\text{res}}=2.5~\mu\text{s},$ and $\nu_{\text{res}}=1.5625~{\text{kHz}}$.
As in \cite{marchese2024,Muppaneni2023,Mattu2024}, the parameters for the channel estimation algorithms are $P_{\max}=15,\ s_{\max}=10,\ \epsilon_{\tau}=10^{-10},\ \epsilon_{\nu}=10^{-2},\ m_{\tau}=n_{\nu}=10$. Finally, as discussed above, the threshold $\epsilon$ for the stopping criterion is set according to the noise variance and OTFS frame parameters. 

\subsection{Simulation Scenario}
The high mobility multipath channel is assumed to have $P$ Rayleigh faded paths with exponential Power Delay Profile (PDP) with decaying time constant of $10~\mu\text{s}$ and maximum delay of $10~\mu\text{s}$. In the considered case study, the propagation paths are close enough so that high IPI occurs. 
The channel is made of $P=4$ propagation paths with delays $\boldsymbol{\tau}=[0~~2.3~~3.15~~7.7]~\mu\text{s}$. The maximum User Equipment (UE) speed is $v_{UE}=190~\frac{\text{m}}{\text{s}} (684 \frac{\text{km}}{\text{h}})$ resulting into a maximum Doppler spread of $\nu_{\max}=\frac{v_{UE}}{c}f_c=3.23~\text{kHz}$, where $c=3\cdot 10^8~\frac{\text{m}}{\text{s}}$ is the speed of light. 
The Doppler shifts of all paths are generated assuming Jakes' Doppler power spectrum (DPS) using $\nu_i=\nu_{\max}\cos(\theta_i)$ where $\theta_i\sim\mathcal{U}[0,2\pi]$. For numerical results, a set of $10^2$ channel realizations has been considered.

Performance metrics are evaluated as functions of the pilot signal-to-noise ratio (PSNR). The pilot SNR is given as
\begin{equation} \label{eq:13}
\text{PSNR}=\frac{E_p}{MN\sigma^2}.
\end{equation}

\begin{figure}
\centering
    \includegraphics[width=0.99\columnwidth]{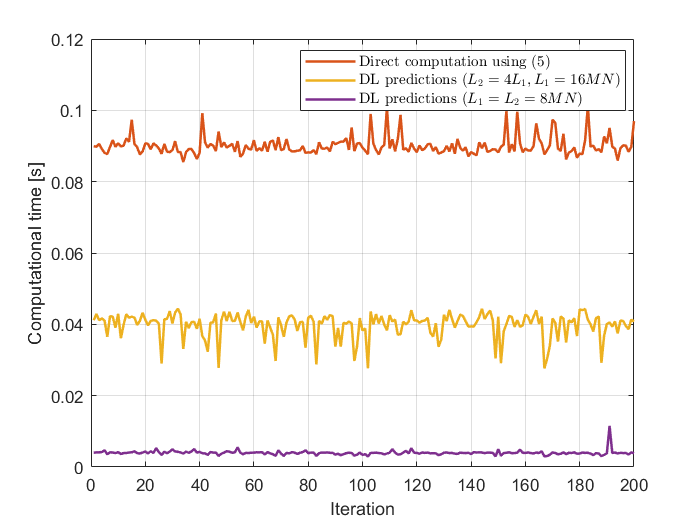}
    \caption{Computational time for different
    inputs.}\label{fig:CompTime}
\end{figure}
\begin{figure}
\centering
    \includegraphics[width=0.99\columnwidth]{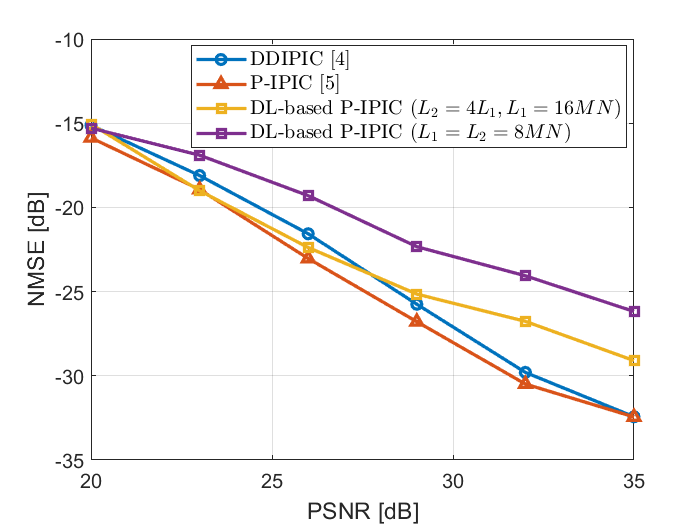}
    \caption{NMSE as function of pilot SNR.}\label{fig:NSMEvsPSNR}
\end{figure}
\begin{figure}
\centering
    \includegraphics[width=0.99\columnwidth]{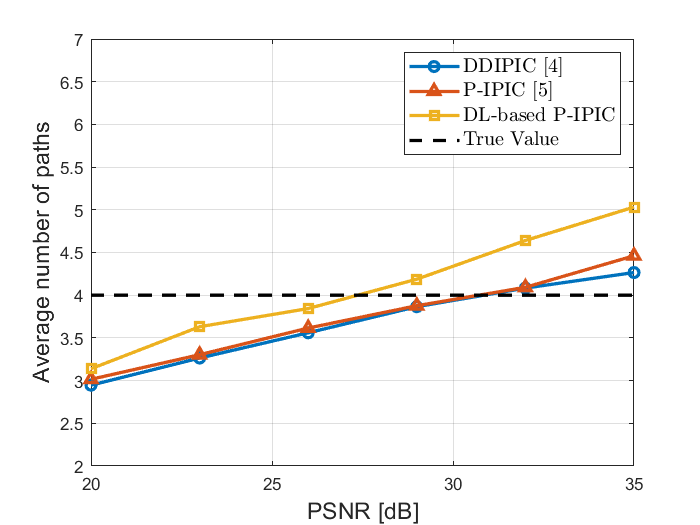}
    \caption{Average number pf paths as function of pilot SNR.}\label{fig:PavgvsPSNR}
\end{figure}

\subsection{Computational Time}
Fig.~\ref{fig:CompTime} shows a comparison in terms of computational time between DL- and model-based approaches. The computational time obtained in simulation is shown for different inputs (200 DD pairs, randomly generated). It can be observed that the FNN-based architecture in Fig.~\ref{fig:FNNarchitecture} effectively enables latency reduction. In particular, in the case $L_1 = L_2 = 8MN$, the computational time is almost reduced by a factor of ten. On the other hand, in the case $L_2 = 4L_1$ with $L_1 = 16MN$, latency increases, as expected. In this case, the computational time is approximately halved (latency reduction of 55\%). It is worth mentioning that the effective latency reduction depends on the practical hardware implementation, meaning that its actual advantage may vary. However, the simulation results demonstrate the potential of the proposed approach.

\subsection{NMSE vs Pilot SNR}
The accuracy of channel estimation algorithms is measured through the Normalized Mean Squared Error (NMSE) computed as
\begin{equation} \label{eq:12}
\text{NMSE}=\mathbb{E}\Bigg[\frac{\|\textbf{H}_{dd}-\hat{\textbf{H}}_{dd}\|^2_F}{\|\textbf{H}_{dd}\|^2_F}\Bigg].
\end{equation} 
Fig.~\ref{fig:NSMEvsPSNR} shows the NMSE as a function of the pilot SNR. Considering the first case ($L_1 = L_2 = 8MN$), it can be observed that a performance loss of $6$ dB is experienced at most, but with a drastically reduced latency. This is due to the limited learning capabilities of the FNNs. On the other hand, when the number of neurons per layer is increased, it can be seen that the DL-based approach offers, for PSNR values ranging from almost $22$ dB to $26$ dB, performance comparable to that of the P-IPIC algorithm. Thus, a performance gain of $1$ dB is achieved compared to DDIPIC. On the contrary, at higher PSNR values, the DL-based approach exhibits a performance loss compared to model-based approaches. This performance loss is at most 3 dB and can be considered acceptable given the advantage in terms of latency achieved using DL predictions. Moreover, this performance reduction can be attributed to the limited capabilities of neural networks to learn the desired relation. In particular, since P-IPIC is based on residuals, it is prone to error propagation, and at higher PSNR values, errors in the estimate lead to residual peaks that are interpreted as small amplitude paths by the algorithm \cite{marchese2024}.

\subsection{Average Number of Paths vs Pilot SNR}
Detection capabilities are investigated looking at the average number of paths detected by the algorithms as a function of the pilot SNR. Simulation results are shown in Fig.~\ref{fig:PavgvsPSNR}. It can be noticed that the DL-based approach ($L_2=4L_1$ with $L_1=16MN$) detects, on average, a higher number of propagation paths compared to model-based approaches. Moreover, as expected, at high PSNR values the DL-based P-IPIC overestimates the number of propagation paths by one. This results is consistent with the performance loss experienced in the NMSE at high PSNR values.

\section{Conclusions and Future Works}
In this work, a DL-based P-IPIC algorithm is proposed to perform low-latency channel estimation for the fractional DD case. Simulation results show that the DL-based P-IPIC can achieve good estimation performance at drastically reduced latency for low to mid PSNR values. In contrast, at higher PSNR values, an acceptable performance loss is unavoidable. 

In future work, a DL-based single-step refinement procedure will be investigated to further reduce the latency and complexity of fractional channel estimation algorithms in high-IPI scenarios.

\section*{Acknowledgment}
This study was partially supported by the European Union under the Italian National Recovery and Resilience Plan (NRRP) of NextGenerationEU, partnership on “Telecommunications of the Future” (PE00000001 - program “RESTART”).

\balance 
\bibliographystyle{IEEEtran}
\bibliography{list}

\end{document}